\documentclass[fleqn,11pt, twocolumn]{wlscirep}
\usepackage[utf8]{inputenc}
\usepackage[T1]{fontenc}
\usepackage{amsmath}
\usepackage{soul}
\usepackage{caption}
\usepackage{float}
\usepackage{graphicx}
\usepackage{subfigure}
\usepackage{epstopdf} 
\usepackage[export]{adjustbox}
\usepackage[justification=justified]{caption}

\usepackage{multirow}
\usepackage[
backend=biber,
style=nature,
sorting=none
]{biblatex}

\usepackage{amsmath}

\title{Manifold Learning for Knowledge Discovery and Intelligent Inverse Design of Photonic Nanostructures: Breaking the Geometric Complexity}
\author[1]{Mohammadreza Zandehshahvar}
\author[1]{Yashar Kiarashi}
\author[1]{Muliang Zhu}
\author[1]{Hossein Maleki}
\author[1]{Tyler Brown}
\author[1, *]{Ali Adibi}

\affil[1]{School of Electrical and Computer Engineering, 
    Georgia Institute of Technology, Atlanta, GA, USA}

\affil[*]{ali.adibi@ece.gatech.edu}

\begin{abstract}

Here, we present a new approach based on manifold learning for knowledge discovery and inverse design with minimal complexity in photonic nanostructures. Our approach builds on studying sub-manifolds of responses of a class of nanostructures with different design complexities in the latent space to obtain valuable insight about the physics of device operation to guide a more intelligent design. In contrast to the current methods for inverse design of photonic nanostructures, which are limited to pre-selected and usually over-complex structures, we show that our method allows evolution from an initial design towards the simplest structure while solving the inverse problem.

\end{abstract}
\begin{document}

\flushbottom
\maketitle

\thispagestyle{empty}


\section{Introduction} \label{Intro}

Photonic nanostructures have been extensively employed for different applications due to their unique features in controlling the spatial, spectral, polarization, and even temporal properties of an optical wavefront with subwavelength feature size. Thanks to recent advances in optical materials and nanofabrication technologies, extensive design flexibility exists for photonic nanostructures through selection of constituent materials and geometrical properties of individual nanostructures (also known as nanoantenna or meta-atoms). This has enabled practical photonic devices for a wide range of applications in computing \cite{shen2017deep,duport2012all}, signal processing \cite{cervera2010reliable}, imaging \cite{colburn2018metasurface}, planar lenses \cite{khorasaninejad2016metalenses, wang2018broadband}, and wireless communication \cite{koenig2013wireless, wang2021towards}, just to name a few. The large number of design parameters (e.g., material selection and geometrical properties of nanoantenna) requires new approaches for inverse design and optimization of these nanostructures. Unfortunately, there has been limited progress in this direction, and the continuous progress in nanofabrication capabilities demands urgent progress in this new direction. In addition, new techniques are needed to find the best device architecture for a given response with minimum device complexity subject to constraints like ease of fabrication, sensitivity to fabrication errors, low optical losses, etc. Such techniques should allow the design algorithm to start from the under-defined architecture and evolve to the most appropriate design beyond what initially envisioned by the designer.   

Considering the large number of design parameters and the ranges of their variations, traditional brute-force inverse design approaches based on exhaustive search \cite{wu2003design} of the design space or evolutionary approaches \cite{bossard2014near} (e.g., genetic algorithms) cannot be used due to the computational complexity of state-of-the-art nanophotonic design problems. More recently, inverse design methods based on artificial intelligence (AI) \cite{ma2018deep,tahersima2019deep,peurifoy2018nanophotonic,chen2020design,ren2020benchmarking,abhishek2020cyclical,ma2020deep,jafar2020tco,deng2020neural,zhelyeznyakov2021deep,nadell2019deep,dinsdale2020deep,du2020expedited,ma2021intelligent,trisno2020applying,unni2020deep,eybposh2020deepcgh,so2020deep,ma2020inverse,kumar2020inverse,singh2020mapping,sheverdin2020photonic,mall2020fast,malkiel2018plasmonic} have shown promising performance in inverse design with considerably reduced computational requirements. However, existing approaches are mainly focused on finding the design parameters of a given nanostructure without changing its structure. In addition, most of the reported techniques only consider finding the design parameters without providing insight about the underlying physics of the device operation, e.g., by providing information about the hidden patterns in the data. More recently, there has been more interest in using AI  to investigate the design-response relation while optimizing the nanostructures \cite{kiarashinejad2020knowledge,kiarashinejad2020deep,hemmatyar2019full,kiarashinejad2019deep,an2020deep,kudyshev2020machine,iten2020discovering,melati2019mapping, kiarashinejad2020geometric,jiang2020deep,wiecha2020deep,  zandehshahvar2020cracking,liu2018training}. New approaches for  “knowledge discovery” in nanophotonics should utilize the “intelligent” aspects of AI rather than using AI primarily as an optimizer.

\begin{figure*}[t]
	\includegraphics[width=\linewidth]{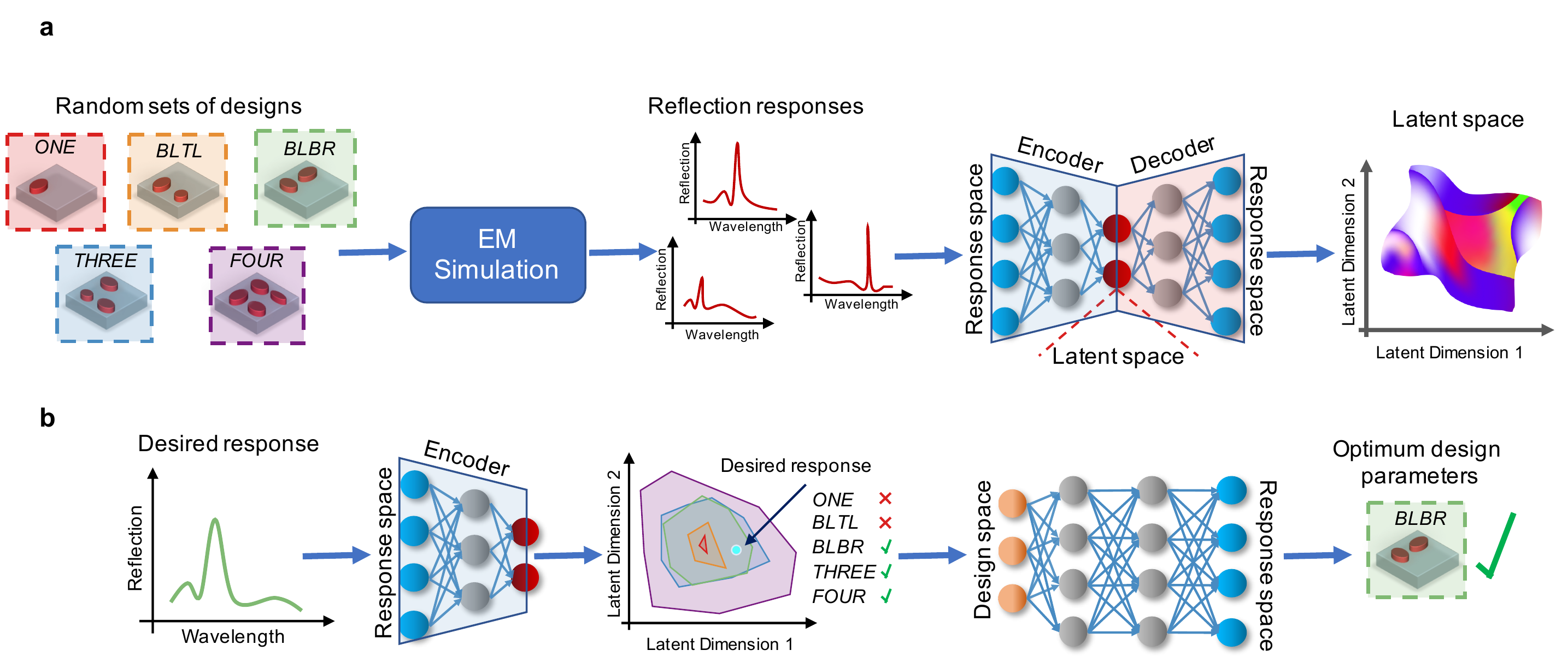} 
	\caption{\textbf{Workflow of the manifold-learning-based design approach}. \textbf{a} Forming the feasible regions and learning sub-manifolds in the latent space. Each sub-manifold corresponds to one of the five nanostructure classes, whose unit cells are shown. Random sets of design parameters are generated for each class, and the corresponding responses are found using an electromagnetic (EM) solver. By training an AE, the dimensionality of the response space is reduced into 2 or 3 and each sub-manifold is modeled using a separate GMM. Each GMM covers the range of feasible responses from a given class of nanostructures. \textbf{b} For inverse design, the dimensionality of the desired response is reduced using the trained AE to observe the feasibility of the response using different classes of nanostructures in \textbf{a}. Using a trained neural network (NN) that relates the design space into the latent response space, we search for the optimum solution with minimal complexity.  It is observed from simulations that the reflection responses have a single resonance peak with a Fano-like lineshape as shown in \textbf{a} and \textbf{b}. } 
	\label{fig: Workflow} 
\end{figure*}
The main focus of this paper is on addressing these two major needs: 1) AI-based techniques to enable inverse design while enabling the evolution of the nanostructure architecture from the initial guess (by the designer) to the most appropriate architecture, and 2) enabling knowledge discovery in photonic nanostructures by uncovering the roles of different design parameters in the final response. For this purpose, we present a new approach based on manifold learning for breaking the geometric complexity of nanophotonic structures during solving the inverse problem. We reduce the dimensionality of the response space by training an autoencoder (AE) \cite{hinton2006reducing} over a set of training data and model the sub-manifolds for different classes of nanostructures with different degrees of design complexity in the latent space (i.e., low-dimensional space) using Gaussian mixture models (GMMs) \cite{murphy2012machine}. This representation also results in understanding the underlying patterns in the data and provides valuable insight about the underlying physics of the nanostuctures. For solving the inverse design, we will map the desired response into the latent space and search over different sub-manifolds with response feasibility to evolve to design candidates with minimal geometric complexity. As a proof of concept and without loss of generality, we apply this method to study dielectric metasurfaces (as a popular class of photonic nanostructures) formed by elliptical hafnium dioxide (HfO$_2$) meta-atoms on a silicon dioxide (SiO$_2$) substrate.

\begin{figure*}[t]
	\includegraphics[width=\linewidth]{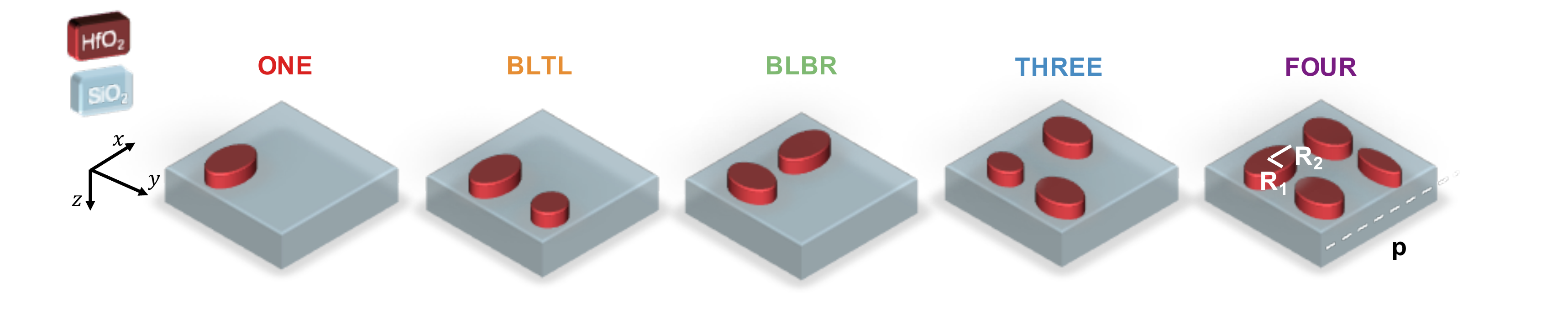} 
	\caption{\textbf{The metasurface unit cells with different geometric complexities}. Each unit cell is one to four HfO$_2$ ellipsoids on a SiO$_2$ substrate. The design parameters are periodicity ($p \in [500,900]$ nm) and radii of the ellipsoids ($R_i \in [45,200]$ nm). The height of all ellipsoids is fixed at $h_{elp}=350$ nm while that of the substrate is assumed to be infinite. The design complexity (i.e., the number of design parameters) of the structures are between 3 and 9 for the simplest (\textit{ONE}) and most complex (\textit{FOUR}) cases, respectively. The resonance features of the selected ellipses result in a single resonance peak in the reflection responses of the metasurface formed by a periodic array of each one of these unit cells (see Fig. \ref{fig: Workflow}). } 
	\label{fig: Structure} 
\end{figure*}

\section{Results}
\subsection{Dielectric metasurfaces with resonant reflection responses}

To show the capabilities of our approach, we study the inverse design of metasurfaces with reflection responses of Fano-like lineshape \cite{hemmatyar2019full} using the unit-cell structures shown in Fig. \ref{fig: Structure}. These structures are composed of a series (one to four) ellipsoids of HfO$_2$ on a SiO$_2$ substrate. The design parameters are periodicity ($ p \in [500,900]$ nm) and the radii of the ellipsoids ($R_i \in [45,200]$ nm). The height of the ellipsoids (350 nm) is fixed due to fabrication limitations, and the substrate is assumed to be infinite in thickness.  The simplest design (\textit{ONE} in Fig. \ref{fig: Structure}) and the most complex one (\textit{FOUR}) have  3 and 9  design parameters, respectively. 

For the AI analysis, a total of 8000 random sets of design parameters for the five unit-cell structures are generated, and the corresponding reflection responses are computed using three-dimensional finite-difference time-domain (3D FDTD) simulations, implemented using Lumerical, in the 300 < $\lambda$ < 850 nm range, where $\lambda$ is the wavelength. The incident beam is a normally-incident plane-wave from the top with linear polarization in the $x$ direction in Figs. \ref{fig: Workflow}a and \ref{fig: Structure}. Each reflection response is sampled at 551 uniformly-placed wavelengths in the operating range. This structure is capable of generating Fano-type reflection responses. Also, when co-polarized resonances of different ellipsoids are strongly coupled in the polarization direction (i.e., $x$-direction in Fig. \ref{fig: Structure}), overall resonant responses with relatively high quality factors (Qs) are observed.

\begin{figure*}[th]
	\includegraphics[width=\linewidth]{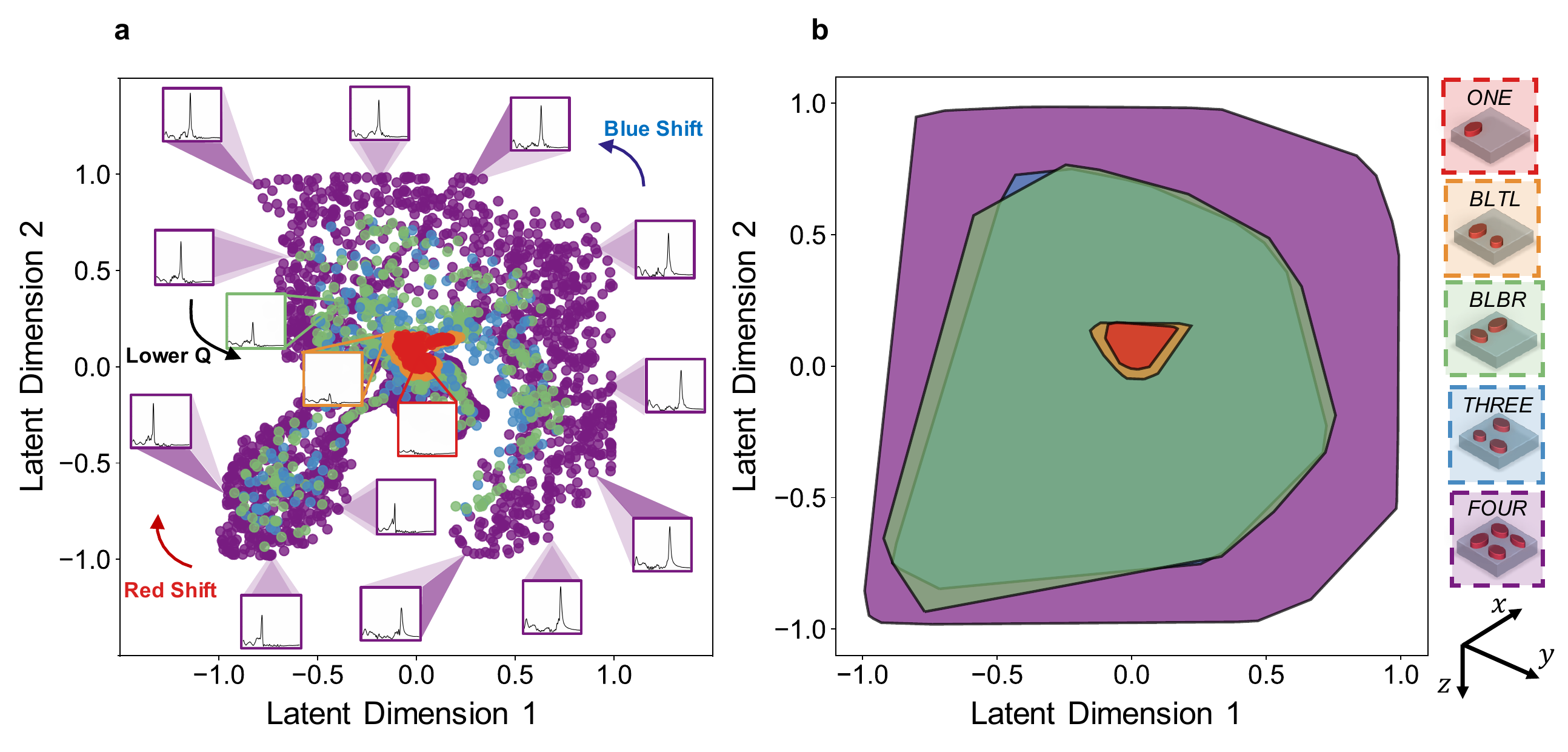} 
	\caption{\textbf{Latent-space representation of the reflection responses}. \textbf{a} Representation of the reflection responses and \textbf{b} the corresponding convex-hulls of the feasible regions for unit-cell structures in Fig. \ref{fig: Structure} in the latent space. The smallest and largest convex-hulls in \textbf{b} correspond to the simplest (\textit{ONE}) and most complex (\textit{FOUR}) structures, respectively. The magnitude of the reflection peak and the Q of the resonance increase as we move from the center of the latent space towards the edges in \textbf{a}. Clockwise (counterclockwise) movement in the latent space results in red (blue) shift in the resonance peaks in the reflection responses.  } 
	\label{fig: Latent} 
\end{figure*}

\subsection{Knowledge discovery using the latent-space representation of the reflection responses}

To form the feasible set of responses for each nanostructure in Fig. 2, we train an AE to reduce the dimensionality of the response space from 550 (i.e., the number of samples in the spectral response) into two or three (i.e., two-dimensional (2D) and three-dimensional (3D) latent spaces, respectively), while minimizing the reconstructed mean-squared error (MSE). In addition, we train a convex-hull using the algorithm explained in Ref. \cite{kiarashinejad2020knowledge} to encompass the range of feasible responses using each of the unit-cell structures in Fig. \ref{fig: Structure}. Figures \ref{fig: Latent}a and \ref{fig: Latent}b show the representation of the responses and the convex-hulls of the feasible regions for each structure in a 2D latent space, respectively. The corresponding results for the 3D latent space are provided in the Supplementary Information. As seen from Fig. \ref{fig: Latent}, the range of feasible responses expands as we increase the design complexity (i.e., the number of ellipsoids in the unit cell). The feasible region of the structure with one ellipsoid (i.e., \textit{ONE}) is the smallest due to both the weak resonance in the nano-antenna and weak coupling between the nano-antennas in the periodic metasurface. An interesting observation from Fig. \ref{fig: Latent}b is the large disparity between the convex-hulls of the \textit{BLTL} and \textit{BLBR} structures while both having two ellipsoids in their unit cell. Also, the convex-hulls of the \textit{BLBR} and \textit{THREE} structures share similar regions despite having different levels of complexity. These non-trivial observations need to be understood using the physics of coupling between different ellipsoids (or meta-atoms).

\begin{figure*}[t]
	\includegraphics[width=\linewidth]{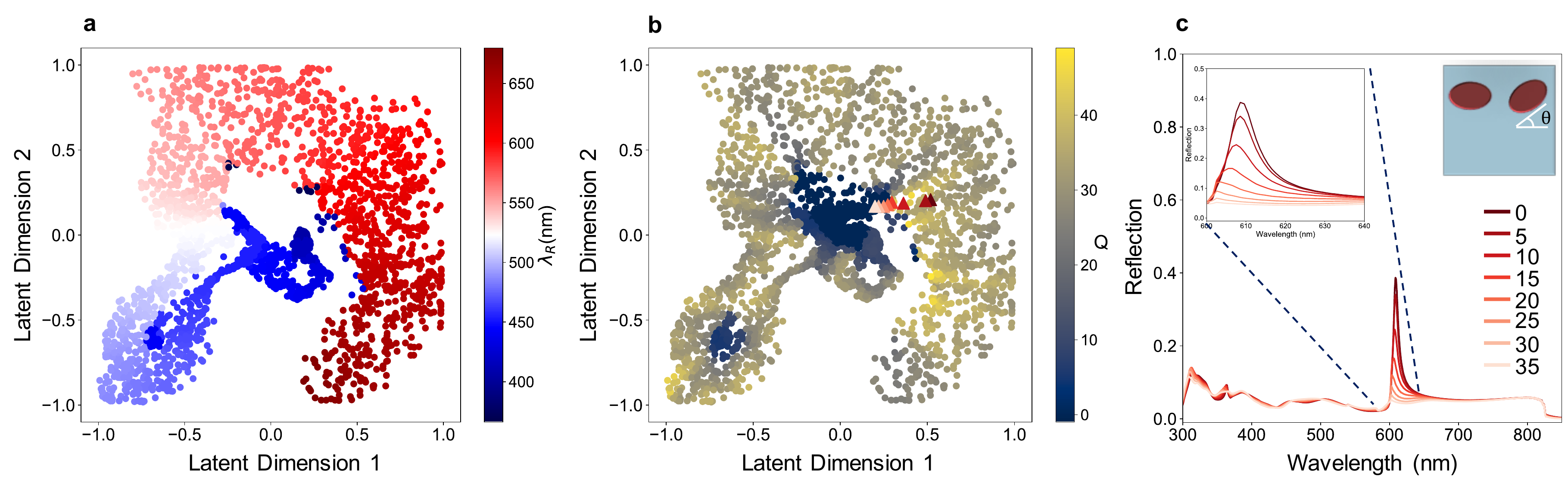} 
	\caption{ \textbf{Q factor and rotation analysis}. \textbf{a} Wavelength distribution of resonances in the latent space. \textbf{b} High/low-Q distribution of the responses in the latent space. \textbf{c} Effect of rotation of the ellipsoid on the reflection response.  } 
	\label{fig: Resonance} 
\end{figure*}

The latent-space representation of the responses (see Fig. \ref{fig: Latent}a) provides priceless information about the underlying patterns in the reflection responses. For example, it is observed that the clockwise movement around the feasible region results in a red shift, and the counter clockwise movement results in a blue shift of the resonances. In addition, Fig. \ref{fig: Latent}a shows an increase in the magnitude and Q of the reflection as we move from the center of the latent space towards the edges. To better quantify the knowledge provided by the manifold-learning approach, we present the color-coded manifolds for the wavelength and the Q of the Fano-type resonances of the reflection resonances in Figs. \ref{fig: Resonance}a and \ref{fig: Resonance}b, respectively. Figure \ref{fig: Resonance}a clearly shows the red (blue) shifts of the resonance wavelength by clockwise (counterclockwise) movements in the latent space, and Fig. \ref{fig: Resonance}b shows that higher Qs are achieved at the corners of the latent space. 

Comparing the feasible responses of structures with different unit cells in Figs \ref{fig: Latent} and \ref{fig: Resonance} suggests that: 1) the \textit{ONE }and \textit{BLTL} structures cannot generate high-Q responses, 2) the \textit{BLBR} structure is far more capable than the \textit{BLTL} structure in forming a variety of different responses despite apparent similarity; 3) The \textit{BLBR} and \textit{THREE} structures have a similar capability in generating high-Q responses despite their different levels of design complexity; and 4) the \textit{FOUR} structure provides the largest range of responses thanks to its highest level of complexity. While some of these conclusions (e.g., 4) might be trivial at the beginning, others (e.g., 2 and 3) are not expected at the first glance. This clearly shows the power of our manifold-learning approach in knowledge discovery in nanophotonics.  

In addition to comparing different structures with different levels of complexities, our manifold-learning approach can provide valuable insight about the roles of different design parameters. To show this capability, we study the effect of rotating one of the ellipsoids in the \textit{BLBR} structure (as the least complex structure with a large range of high-Q responses) on the reflection response while keeping other design parameters fixed (see Fig. \ref{fig: Resonance}c). It is clear from Fig. \ref{fig: Resonance}c that by rotating one of the ellipsoids (i.e., increasing $\theta$ from 0), both the peak reflection magnitude and the Q decrease with a minor resonance wavelength shift, however, the reflection response outside the resonance range stays almost the same. To see this in the latent space, the corresponding responses, after dimensionality reduction, have been shown in Fig. \ref{fig: Resonance}b using triangles with the same colors as those of the actual responses in Fig. \ref{fig: Resonance}c. The movement of these triangles towards the center of the latent space and the lack of considerable clockwise or counter clockwise rotation by increasing $\theta$ in Fig. \ref{fig: Resonance}b confirms the ability of the manifold-learning approach in uncovering the observed role of $\theta$. 

The amount of visually observable information about different classes of unit cells, seen from Figs. \ref{fig: Latent} and \ref{fig: Resonance} shows the efficacy of our manifold-learning approach in knowledge discovery, i.e., providing valuable observable insight about the physics of nanophotonic device operation. 

\subsection{Inverse design using the manifold-learning approach }

To better quantify the effectiveness of each unit-cell structure in Fig. \ref{fig: Structure} in forming a desired response, we model the sub-manifolds of the corresponding responses in the latent space for each structure using GMMs (see Methods and Supplementary Information). These GMMs provide the levels of feasibility of achieving any given reflection response for metasurfaces with different unit-cell structures in Fig. \ref{fig: Structure}, which will be helpful in the inverse design. 

\begin{figure*}[h!]
	\includegraphics[width=\linewidth]{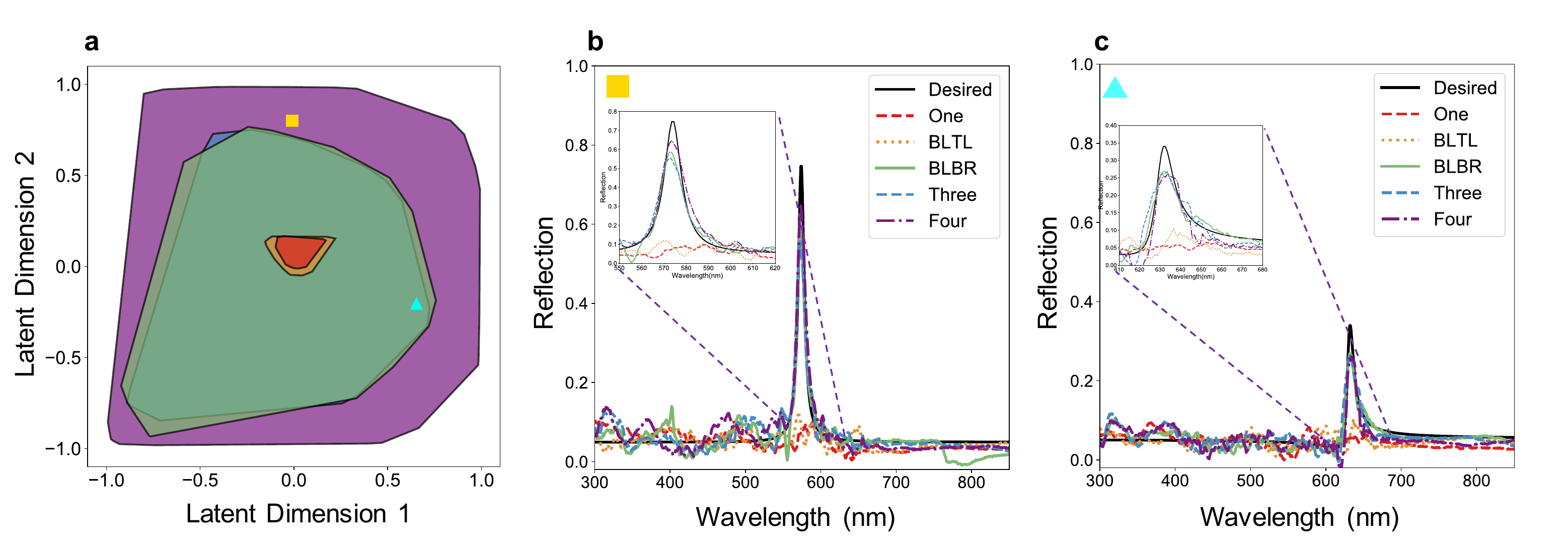} 
	\caption{  \textbf{Inverse design of Fano reflection responses}. \textbf{a} Representation of two desired responses with an ideal resonance lineshpe (zero reflection outside the resonant region) in the latent space.  \textbf{b-c} Desired (ideal) and the corresponding optimized responses (found by our inverse design approach) for different unit-cell structures in Fig. \ref{fig: Structure}. } 
	\label{fig: Design} 
\end{figure*}

To find a structure that generates a desired reflection response, the first step is to find the corresponding point in the latent space by reducing the dimensionality of the desired response using the trained AE (see the two examples in Fig. \ref{fig: Design}a). Next, we find the log-likelihood of the feasibility for the desired response using the five different unit cells in Fig. \ref{fig: Structure} by employing their corresponding GMMs. We select the unit-cell structure with the highest log-likelihoods. Once the unit-cell of the metasurface is selected, we use exhaustive search with a separately trained neural network (see Methods) for the forward problem (i.e., connecting the design and response spaces) of that metasurface to find the optimum design parameters. 
\begin{table*}[h!]
  \caption{The optimal design parameters (in nm), normalized MSE (NMSE), negative log-likelihood, for the Fano reflection response in Fig. \ref{fig: Design}b}
  \label{table: table-1}
  \centering
  \tabcolsep=0.11cm
  \begin{tabular}{lllllllllllll}
    \toprule
    \multicolumn{12}{c}{Design Parameters}                 \\
    \cmidrule(r){2-10}
    Structure     & p  &R1BL &R2BL &R1BR &R2BR &R1TL &R2TL &R1TR &R2TR &NMSE &$-\log(p)$ \\
    \midrule
    One    &783 &178 &17&0&0&0&0&0&0&0.60&186.74     \\
    BLTL     &599  &63 &63 &0&0 &173 &130 &0&0&0.546&468.08     \\
    BLBR     &758      &149 &104 &148 &121 &0&0&0&0&0.095&3.01  \\
    THREE     &684       &151 &135 &170 &132 &160 &86 &0&0 &0.084&3.24 \\
    FOUR    &769      &123 &74 &148 &99 &142 &80 &128 &148 &\textbf{0.083}&2.18 \\
    \bottomrule
  \end{tabular}
\end{table*}
\begin{table*}[h!]
  \caption{The optimal design parameters (in nm), NMSE, negative log-likelihood, for the Fano reflection response in Fig. \ref{fig: Design}c}
  \label{table: table-2}
  \centering
  \tabcolsep=0.11cm
  \begin{tabular}{lllllllllllll}
    \toprule
    \multicolumn{12}{c}{Design Parameters}                 \\
    \cmidrule(r){2-10}
    Structure     & p  &R1BL &R2BL &R1BR &R2BR &R1TL &R2TL &R1TR &R2TR &NMSE &$-\log(p)$ \\
    \midrule
    One    &672 &78 &180&0&0&0&0&0&0&0.36&276.8     \\
    BLTL     &777  &76 &64 &0&0 &168 &157 &0&0&0.31& 214.26     \\
    BLBR     &851      &143 &134 &163 &132 &0&0&0&0&\textbf{0.056}&0.57  \\
    THREE     &833       &172 &92 &175 &175 &76 &72 &0&0 &0.086&1.17 \\
    FOUR    &7846     &170 &150 &142 &134 &153 &113 &78 &78 &0.099&1.24 \\
    \bottomrule
  \end{tabular}
\end{table*}
Figure \ref{fig: Design} shows the implementation of the inverse design problem for two desired responses with high and moderate Qs (see Figs.  \ref{fig: Design}b and  \ref{fig: Design}c, respectively). To compare the effectiveness of metasurfaces with different unit cells and the importance of the GMMs, we implement each design using all possible unit cells (regardless of the design feasibility), and the corresponding results are shown in Figs  \ref{fig: Design}b and  \ref{fig: Design}c as well as Tables \ref{table: table-1} and \ref{table: table-2}, respectively. This experiment mimics the conventional design approaches focused on finding the design parameters regardless of the feasibility of the response. Figure \ref{fig: Design}a suggests that the desired response with Q = 52 can only be generated using the \textit{FOUR} structure. This is confirmed by comparing the actual optimal responses (see Fig. \ref{fig: Design}b) and the negative log-likelihood values ($-\log (p)$ in Table \ref{table: table-1}). Similarly, Fig. \ref{fig: Design}a suggests that the response with Q = 42 can be generated by \textit{BLBR}, \textit{THREE}, and \textit{FOUR} structures. This is confirmed by different optimal responses and log-likelihoods from Fig. \ref{fig: Design}c and Table \ref{table: table-2}, respectively. Tables \ref{table: table-1} and \ref{table: table-2} also provide means for using a trade-off between design complexity and the response errors. The importance of the manifold-learning approach is that it enables the consideration of the feasibility before attempting to design a device using a pre-selected structure.

\section{Discussion}
Of the four insights and conclusions from Figs \ref{fig: Latent} and \ref{fig: Resonance} (as discussed in Section 2.2), the difference in the size of the convex-hulls of \textit{BLTL} and \textit{BLBR} structures and the similarity of the convex-hulls of \textit{BLBR} and \textit{THREE} structures contradict the expected increase of the range of feasible responses and the design complexity. To explain the reason for these important conclusions from the manifold-learning approach, we perform the near-field analysis of these three structures using the 3D FDTD technique.  

\begin{figure*}[t]
\centering
	\includegraphics[width=0.6\linewidth]{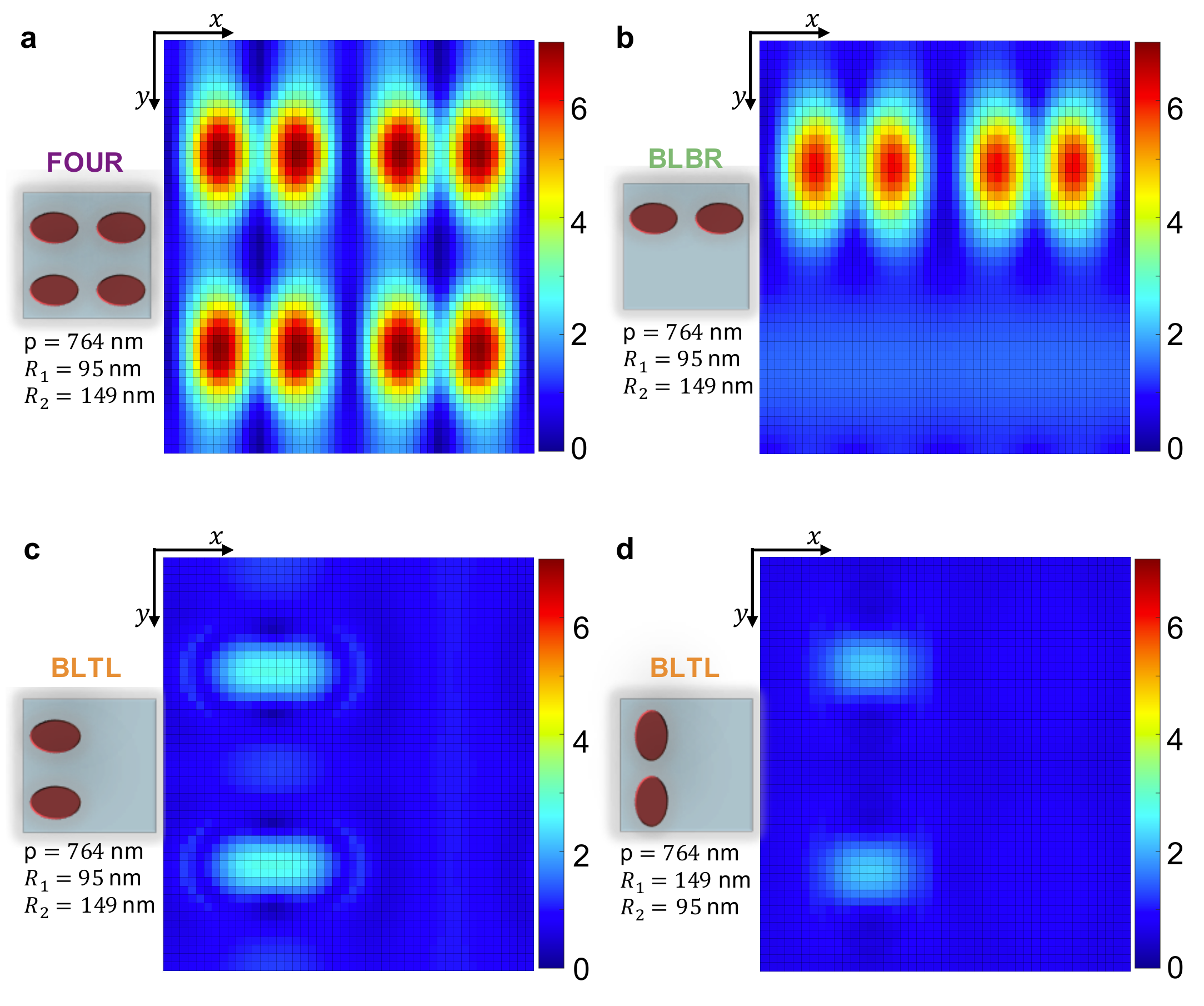} 
	\caption{ \textbf{Nearfield enhancement}. The nearfield simulation of the \textbf{a} \textit{FOUR}, \textbf{b}, \textit{BLBR}, and \textbf{c}, \textbf{d} \textit{BLTL} structures under normal illumination with a uniform planewave at $\lambda$ = 576 nm, 567 nm, 390 nm, and 383 nm, respectively, with linear polarization in the $x$-direction. The color code shows the electric-field amplitude divided by the amplitude of the incident filed (as a measure of field enhancement by the resonant structures).} 
	\label{fig: Nearfield} 
\end{figure*}
Figure \ref{fig: Nearfield} shows the resonant nearfield enhancement by the \textit{FOUR}, \textit{BLBR}, and \textit{BLTL} structures. Due to the resonance of the individual ellipsoids at the incident wavelength, the field enhancement near these structures is expected. An important observation is the contrast between field enhancements in the $x$- and $y$- directions for all structures in Fig. \ref{fig: Nearfield}. This contrast is caused by the different levels of coupling between individual (resonant) ellipsoids in the directions parallel and perpendicular to the incident polarization (i.e., $x$- and $y$- directions, respectively). Our analysis shows that the maximum field enhancement (or the strongest light-matter interaction) is obtained for ellipsoids with resonances co-polarized with the incident polarization with strong coupling in the direction of incident polarization (i.e., $x$-direction in Fig. \ref{fig: Nearfield}). It is clear that the coupled ellipsoids in the x-direction in both \textit{FOUR} and \textit{BLBR} structures (Figs. \ref{fig: Nearfield}a and \ref{fig: Nearfield}b, respectively) are responsible for the strong field enhancement while the coupled ellipsoids in the $y$-direction in the \textit{BLTL} structure (Fig. \ref{fig: Nearfield}c) cannot provide such enhancement. As a result, the resonance strength and Q of reflection in \textit{FOUR} and \textit{BLBR} structures are somehow similar and they both are much stronger than those of the \textit{BLTL} structure as observed from Figs. \ref{fig: Latent} and \ref{fig: Resonance}. 

It is important to note that the different distances between adjacent ellipsoids in the $x$- and $y$-directions are not the main contributors to the difference of the light-matter interaction strength in different structures. To resolve this potential confusion, Fig. \ref{fig: Nearfield}d shows a different structure that includes two ellipsoids with small spacing in the $y$-direction. It is clear from Fig. \ref{fig: Nearfield}d that despite this closeness, the orthogonality of the coupling direction and the incident polarization results in weak field enhancement. A further evidence of this fact is seen from Fig. \ref{fig: Resonance}c, where by rotating one ellipsoids in the \textit{BLBR} structure (and thus, weakening the coupling in the polarization direction), the resonance strength is reduced and the responses move towards the center of the latent space. The simulation results for the field enhancement in a variety of structures with different levels of coupling in $x$- and $y$-directions are provided in the Supplementary Information to further clarify this insight. Nevertheless, this discussion clearly shows the capability of our manifold-learning approach in uncovering the physics of device operation, which can be used to form more effective designs. For example, by using this insight, the unit-cell structures with coupling perpendicular to the incident field will be excluded from the design options before any design attempt. More importantly, this insight excludes the option of using rotated ellipsoids for the design, which considerably reduces the computational requirements for any inverse-design approach. 

In addition to the valuable insight about the device operation physics, the manifold-learning approach helps forming more intelligent designs. First, it shows the level of feasibility of a response using a given class of structures. Secondly, it provides a series of options with different levels of complexity (and potentially robustness, although not discussed in this paper) for the design. As an example, our approach provides three options for the design problem in Fig. \ref{fig: Design}c. It allows moving from a more complex structure (e.g., \textit{FOUR} in this case) to the least complex one (\textit{BLBR}) in our simple approach of using five options for the design. For a more complex design tool, we envision using a trained algorithm with considerably more options beyond what an average designer can consider. Such a design tool will allow the evolution of an initial design structure by the user to a different final design that might be considerably less complex, more robust, less power-hungry, etc. Considering the fact that in many cases, the initial designs are motivated more by using the utmost fabrication capabilities (and thus, most complex structures), such evolutionary design approaches will be very helpful in forming practical structures with optimal resource management.

\section{Methods}
\subsection{Electromagnetic simulations}
The 3D FDTD simulations are conducted with the commercial software Lumerical. The simulation domain is limited to one period (p) in the lateral directions (i.e., $x$ and $y$ in Fig. \ref{fig: Structure}) and perfectly matched layers are used on the top and bottom layers (in the $z$-direction in Fig. \ref{fig: Structure}) due to the periodicity of the structures. 

\subsection{Manifold learning}

To form the latent space of the responses, an AE is trained on a total of 6000 reflection responses obtained by 3D FDTD simulations for the random sets of design parameters of the structures in Fig. \ref{fig: Structure}. Each reflection response is calculated in the 350 nm <  $\lambda$ < 800 nm, and it is sampled uniformly with 550 samples in this range. The dimensionality of the reflection responses is reduced from 550 into 2 and 3 using the trained AE with 11 layers (550, 200, 100, 50, 20, d, 20, 50, 100, 200, 550 nodes at each layer, respectively, where d is the dimension of the latent space). The hidden layers have tangent-hyperbolic (tanh(.)) activation functions, and the input and output layers have linear activation functions. The MSE loss is minimized during the training using Adam optimizer in Python. The training is stopped after 500 epochs if the required MSE is not reached. 

GMMs are used for modeling the sub-manifold of the responses in the latent space for each design complexity. The distance metric is set to correlation, and the maximum distance is 0.3, with a maximum of 5 Gaussian distributions for each model. The GMMs are trained on the training samples in 2D and 3D spaces. 

\subsection{Inverse Design}
To find the optimum design parameters, a feed-forward NN is trained from the design to the response space. The network has 8 layers with 9, 20, 50, 100, 100, 200, 400, 500 nodes at each layer. The activation functions of the hidden layers is tanh(.). The design parameters are normalized to have zero mean and unit standard deviation. The weights of the NN are trained using the Adam optimizer in Python to minimize the MSE. 

To perform the inverse design with a desired reflection response and a given design complexity (i.e., a given unit-cell structure in Fig. \ref{fig: Structure}), we use the exhaustive search (with $10^6$ random sets of design parameters) in the design space using the trained feed-forward NN. The optimum solution with minimum mean-squared distance to the desired reflection response for each design complexity is reported as the solution.

Note that this is the simplest approach for the inverse design using AI. In a more aggressive approach with less computation requirements, recent techniques like training a pseudo-encoder and combining an inverse AI design (from the response space to the reduced design space) with a considerably smaller exhaustive search (from the design space to the reduced design space), as explained in Ref. \cite{kiarashinejad2020deep} can be used.

All of the AI algorithms are implemented in Python and Keras on a system with Core i7 CPU, one RTX2080 GPU, and 32 GB of RAM.



\section*{Competing interests}
The Authors declare no Competing Financial or Non-Financial Interests. 

\section*{Acknowledgment}
The work was funded by Defense Advanced Research Projects Agency (D19AC00001, Dr. R. Chandrasekar).








\printbibliography

\end{document}